# Leveraging Blockchain and Proxy Re-Encryption to secure Medical IoT Records

Abdou-Essamad Jabri[1,2][0009-0003-3510-0087], Cyril Drocourt[2][0000-0003-1636-9462],

Mostafa Azizi[1][0000-0001-9936-3001] and Gil Utard[2][0000-0003-3081-6185]

[1] MATSI Laboratory, Mohammed First University, Oujda, Morocco
[2] MIS Laboratory, University of Picardie Jules Verne, Amiens, France
{abdou-essamad.jabri, azizi.mos}@ump.ac.ma
{cyril.drocourt,gil.utard}@u-picardie.fr

**Abstract.** The integration of the Internet of Things (IoT) in healthcare has revolutionized patient monitoring and data collection, allowing real-time tracking of vital signs, remote diagnostics, and automated medical responses. However, the transmission and storage of sensitive medical data introduce significant security and privacy challenges. To address these concerns, blockchain technology provides a decentralized and immutable ledger that ensures data integrity, , and transparency. Unlike public blockchains, private blockchains are permissioned; the access is granted only to authorized participants; they are more suitable for handling confidential healthcare data. Although blockchain ensures security and trust, it lacks built-in mechanisms to support flexible and controlled data sharing; This is where Proxy Re-Encryption (PRE) comes into play. PRE is a cryptographic technique that allows encrypted data to be re-encrypted for a new recipient without exposing it to intermediaries. We propose an architecture integrating private blockchain and PRE to enable secure, traceable, and privacy-preserving data sharing in IoT-based healthcare systems. Blockchain guarantees tamper-proof record-keeping, while PRE enables fine-grained access control, allowing medical professionals to securely share patient data without compromising confidentiality. This combination creates a robust security framework that enhances trust and efficiency in digital healthcare ecosystems.

## 1 Introduction

The Internet of Things (IoT) in healthcare refers to connected medical devices and sensors that collect, transmit, and analyze patient data in real time. This technology improves patient monitoring, early disease detection, and overall healthcare efficiency. However, it also introduces security and privacy risks due to sensitive health data being transmitted over networks. Blockchain is a decentralized and immutable ledger that enhances security, transparency, and trust in data transactions. In healthcare, it can se-curely store and share patient records, ensuring data integrity and reducing fraud. Private blockchains, unlike public ones, are restricted to authorized users, making them more suitable for sensitive healthcare applications. These blockchains allow controlled access while maintaining decentralization benefits. Proxy Re-Encryption (PRE) is a cryptographic technique that enables a third party (proxy) to transform encrypted data from one key to another without decrypting it. In healthcare, PRE allows secure data sharing between medical professionals while ensuring that unauthorized users cannot access the information. We propose an architecture that combines private blockchain and PRE to enhance data security and access control in IoT-based healthcare systems. The blockchain ensures data integrity, traceability, and decentralized storage, while PRE allows secure and flexible data sharing among authorized users. This integration addresses privacy concerns by enabling fine-grained access control without exposing patient data to intermediaries.

## 2 Related Work

### 2.1 The Internet of Things (IoT) in Healthcare

IoT has created a large network of interconnected devices that collect and transmit data autonomously[1]. In healthcare, IoT devices such as wearable fitness trackers, smart blood pressure monitors, and implantable devices (e.g., pacemakers) continuously generate critical patient data. These devices enable:

- Remote Monitoring: Patients' vital signs can be monitored in real-time, reducing the need for frequent hospital visits.
- Early Detection: Continuous data collection allows early detection of abnormalities, enabling timely medical intervention.
- Personalized Medicine: Data-driven insights from IoT devices support tailored treatment plans based on individual patient needs [2].

However, IoT systems often rely on centralized cloud servers for data storage and processing, which introduces vulnerabilities such as single points of failure, data breaches, and lack of transparency. By integrating IoT with blockchain, these challenges can be addressed. Blockchain ensures that patient data remains secure, traceable, and tamper-proof while giving patients greater control over their information.

### 2.2 Blockchain and Smart Contracts in IoT-Enabled Healthcare

In IoT-enabled healthcare systems, blockchain provides a secure infrastructure for managing the flow of sensitive data. To enhance its capabilities, smart contracts are integrated into the blockchain. Smart contracts are self-executing programs stored on the blockchain that automatically enforce predefined rules and conditions. For example, in a healthcare setting, a smart contract can ensure that only authorized users, such as doctors or researchers, can access patient data, and only under specific conditions (e.g., patient consent).

Smart contracts play a pivotal role in this work by embedding the Proxy Re-Encryption (PRE) process. They:

1. Automate the secure sharing of patient data by ensuring that re-encryption operations occur only when authorized.
2. Manage re-encryption keys without exposing them to unauthorized parties, ensuring confidentiality.
3. Provide an auditable and tamper-proof record of access and re-encryption operations, leveraging the immutability of the blockchain.

### 2.3   Proxy Re-Encryption

While blockchain ensures security and transparency, it does not directly address the need for fine-grained access control over encrypted data. This is where Proxy Re-Encryption (PRE) comes into play. PRE is a cryptographic technique that allows encrypted data to be securely shared with authorized users without exposing the original encryption key [3].

This article proposes an architecture that combines IoT, blockchain, and PRE to create a secure and efficient framework for sharing sensitive healthcare data. By leveraging the strengths of private blockchain and cryptographic techniques, the proposed approach addresses critical challenges in IoT-enabled healthcare, including data privacy, access control, and accountability.

### 2.4   Proxy Re-Encryption with blockchain

[4] introduced a Proxy Re-encryption Enabled Secure and Anonymous IoT Data Sharing Platform Based on Blockchain. [5] also proposed a Proxy Re-Encryption Approach to Secure Data Sharing in the Internet of Things Based on Blockchain, emphasizing the importance of data confidentiality, integrity, and security in cloud environments. [6] discussed an Approach Towards Secure Sharing of Healthcare Records with Blockchain, highlighting the need for patient privacy and data security in sharing sensitive information with healthcare facilities. [7] presented a Blockchain Framework for Secured On-Demand Patient Health Records Sharing, focusing on giving patients full access and control over their health records. [8] proposed a Trustworthy IoT Data Streaming Using Blockchain and IPFS, providing a solution for resource-constrained IoT streaming devices to transfer data chunks in a decentralized, transparent, and secure manner. [9] introduced a Secure Data Sharing for Cross-domain Industrial IoT Based on Consortium Blockchain, focusing on constructing trust among different domains in IIoT. [10] presented a Blockchain-Based Local-Global Consensus Manager to enable a secure IoT environment and enhance security in IoT networks. Lastly, [11] discussed a Dual Blockchain-based Data Sharing Mechanism with Privacy Protection for Medical Internet of Things, emphasizing the importance of privacy protection in sharing medical IoT data.

## 3 IoT healthcare architectures with Blockchain

To provide a meaningful comparison, we review three notable architectures that also integrate blockchain and PRE for secure data sharing:

### 3.1 Decentralized Healthcare Data Sharing Using Blockchain and PRE

This architecture [12] focuses on decentralized healthcare data sharing by integrating blockchain and Proxy Re-Encryption (PRE). It eliminates the need for a central authority by using a **public or consortium blockchain**, ensuring trust, transparency, and secure data delegation. Instead of storing actual patient data on-chain, the blockchain holds metadata such as delegation policies and re-encryption transaction logs. Patient data is stored off-chain in cloud databases, and PRE allows patients to generate **re-encryption keys** that enable authorized users, such as doctors or researchers, to securely access data.

While this architecture **enhances trust and transparency**, it has notable limitations. Using a **public blockchain** means metadata exposure can lead to privacy concerns, as access patterns and transaction logs are visible. Additionally, the lack of **smart contracts** means access control and re-encryption must be handled manually, increasing the risk of human error. Moreover, **scalability issues** arise due to the high computational overhead and latency associated with public blockchain consensus mechanisms.

### 3.2 Blockchain-PRE Systems for IoT

This architecture [13] applies blockchain and Proxy Re-Encryption (PRE) to secure data sharing in IoT environments. It is designed to enhance transparency and prevent unauthorized access to medical data collected from IoT devices. The system consists of three main components: a public blockchain for recording metadata (such as access logs and delegation records), IoT devices for data collection, and PRE for secure delegation. IoT devices encrypt medical data using public-key cryptography, and the encrypted data is stored off-chain in a cloud or distributed file system. The PRE mechanism ensures that authorized users can decrypt data without exposing the original encryption key.

While this approach improves security and decentralization, it suffers from significant scalability issues. Public blockchain consensus mechanisms, such as Proof of Work, introduce high computational overhead, making real-time IoT applications inefficient. Moreover, the lack of automation in re-encryption and access control means that manual intervention or proxy-based enforcement is required, which reduces system efficiency. Additionally, public blockchain use leads to metadata exposure, which may compromise privacy by making delegation logs visible to all participants.

### 3.3 Attribute-Based PRE with Blockchain

This architecture [14] enhances security and access control by **combining blockchain, Proxy Re-Encryption (PRE), and Attribute-Based Encryption (ABE)**. ABE allows data owners to define **attribute-based access policies**, ensuring that only users with specific attributes (e.g., "doctor" or "researcher") can decrypt sensitive medical data. The blockchain (public or consortium-based) serves as a **secure ledger** for storing metadata, delegation records, and access policies, ensuring immutability and trust. PRE further enhances security by enabling patients to generate re-encryption keys for specific users, ensuring **fine-grained control over data access**.

Despite its strong **privacy and security features**, this architecture has **significant computational costs** due to ABE, making it **less suitable for IoT devices** with limited processing power. Additionally, **managing dynamic attributes** (e.g., updating user roles or policies) is complex and time-consuming. Another drawback is the **lack of automation**, as smart contracts are not integrated, requiring manual enforcement of access policies and re-encryption operations.

## 4 Our Proposed Architecture

Our proposed architecture is designed to enhance privacy, scalability, and automation in IoT-enabled healthcare systems by integrating IoT devices, a private blockchain, smart contracts, and Proxy Re-Encryption (PRE). IoT medical devices, such as wearables and implanted sensors, collect and encrypt patient data using the patient's public key before transmitting it securely. Instead of using a public blockchain, which exposes metadata and suffers from scalability issues, our solution employs a private blockchain ensuring that only authorized actors can access the ledger, and thus improving privacy.

A key innovation in our approach is the use of smart contracts to automate the re-encryption process. When a patient wants to share encrypted medical data, they generate a re-encryption key, allowing a specific recipient (such as a doctor) to securely access the data. The smart contract enforces access control, ensuring that even the proxy performing the re-encryption cannot see the plaintext data. Additionally, by incorporating Attribute-Based Encryption (ABE) alongside with PRE, our architecture enables fine-grained access control, allowing users with specific attributes (e.g., "cardiologist" or "researcher") to decrypt data, while still maintaining the ability to delegate access securely.

This architecture significantly improves the whole security upon previous ones. Unlike public blockchain-based systems, which suffer from high latency and metadata exposure, our private blockchain ensures faster transaction processing and better scalability for IoT applications. Furthermore, the automation of access control through smart contracts removes the need for manual enforcement, reducing human errors and increasing system efficiency. By integrating together PRE, ABE, and smart contracts, our system achieves a relevant balance between security, efficiency, and privacy, making it a superior choice for real-world healthcare environments.

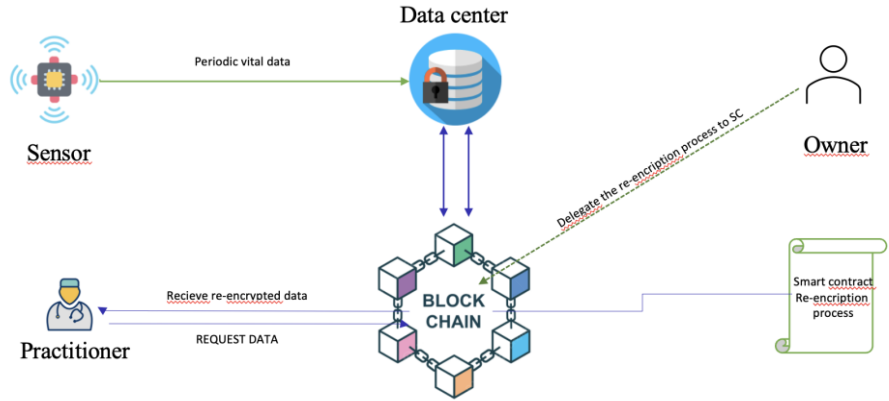

**Fig. 1.** Our proposed architecture

Table 1 compares and summarizes the differences between our architecture and the existing ones.

**Table 1.** Comparative Analysis of Blockchain and Proxy Re-Encryption Architectures for Secure Healthcare Data Sharing

| Architecture | Decentralized Healthcare Data Sharing [x] | Blockchain-PRE for IoT [x] | Attribute-Based PRE with Blockchain [x] | Our Proposed Architecture |
|---|---|---|---|---|
| **Blockchain Type** | Public/Consortium | Public | Public/Consortium | Private |
| **Use of (PRE)** | Yes | Yes | Yes | Yes |
| **Use of (ABE)** | No | No | Yes | Yes |
| **Use of Smart Contracts** | No | No | No | Yes |
| **Data Storage** | Off-chain (cloud/local DBs) | Off-chain (cloud/local DBs) | Off-chain (cloud/local DBs) | Off-chain (cloud/local DBs) |
| **Metadata Storage on Blockchain** | Delegation policies & access logs | Delegation policies & access logs | Delegation policies & access logs | Re-encryption transactions & access logs |
| **Privacy Protection** | Moderate (metadata exposure on public chains) | Moderate (metadata exposure on public chains) | High (ABE restricts access) | High (private blockchain + smart contracts) |
| **Automation of Re-Encryption** | No (manual operations) | No (manual operations) | No (manual operations) | Yes (via smart contracts) |
| **Scalability for IoT** | Limited (public blockchain overhead) | Limited (mining delays) | Moderate (ABE computational costs) | High (optimized for IoT) |
| **Security Level** | High (blockchain & PRE) | High (blockchain & PRE) | Very High (ABE + PRE) | Very High (Smart Contracts + PRE + ABE) |
| **Key Advantage** | Decentralized trust, transparency | Secure IoT data sharing | Fine-grained access control | Efficient, automated, and scalable IoT security |
| **Key Limitation** | No automation, metadata exposure | No automation, IoT scalability issues | High computational complexity | Slightly higher complexity due to smart contracts |

## 5  Conclusion

Blockchain ensures security by preventing data tampering, while Proxy Re-Encryption (PRE) enhances privacy by enabling controlled data sharing. Our architecture combining these two technologies solves well the security and efficiency challenges, but it introduces unfortunately new complexities that require further optimization. The integration of smart contracts with PRE and ABE, requires a careful system design to ensure efficiency and prevent vulnerabilities in the automation process. Additionally, private blockchain networks require administrative oversight, meaning that while they provide enhanced security and control, they also necessitate trusted governance. Finally, although our system is more scalable for IoT than public blockchain solutions, further optimizations are needed to minimize energy consumption and computational costs for resource-limited IoT devices.

Despite these challenges, our proposed system represents a major step forward in secure healthcare data sharing, offering a practical and privacy-preserving solution for real-world applications. Future work can focus on enhancing interoperability with other healthcare systems, optimizing smart contract efficiency, and exploring hybrid blockchain models to balance decentralization and performance.

8# 6    References

1. Islam, S. M. R., Kwak, D., Kabir, M. H., Hossain, M., & Kwak, K. S. (2015). The internet of things for health care: A comprehensive survey. *IEEE Access, 3*, 678-708.
2. Zhang P, White J, Schmidt DC, Lenz G, Rosenbloom ST. FHIRChain: Applying Blockchain to Securely and Scalably Share Clinical Data. Comput Struct Biotechnol J. 2018 Jul 29;16:267-278.
3. Blaze, M., Bleumer, G., & Strauss, M. (1998). Divertible protocols and atomic proxy re-encryption. *Proceedings of the 21st Annual International Conference on Cryptology*, 127-143.
4. Ahsan Manzoor; An Braeken; Salil S. Kanhere; Mika Ylianttila; Madhsanka Liyanage; "Proxy Re-encryption Enabled Secure and Anonymous IoT Data Sharing Platform Based on Blockchain",   J. NETW. COMPUT. APPL.,  2021.
5. Kwame Opuni-Boachie Obour Agyekum; Qi Xia; Emmanuel Boateng Sifah; Christian Nii Aflah Cobblah; Hu Xia; Jianbin Gao;  "A Proxy Re-Encryption Approach to Secure Data Sharing in The Internet of Things Based on Blockchain",   IEEE SYSTEMS JOURNAL,  2021.
6. Phapale Pankaj; Satpute Sachin; Wakchaure Shivam; Wale Shubham;  "An Approach Towards Secure Sharing of Healthcare Records with Blockchain",   INTERNATIONAL JOURNAL OF ADVANCED RESEARCH IN SCIENCE, ...,  2021.
7. Meryem Abouali; K. Sharma; O. Ajayi; T. Saadawi;  "Blockchain Framework for Secured On-Demand Patient Health Records Sharing",   2021 IEEE 12TH ANNUAL UBIQUITOUS COMPUTING, ELECTRONICS & ...,  2021.
8. Haya R. Hasan; K. Salah; Ibrar Yaqoob; Raja Jayaraman; Saša Pesic; Mohammed A. Omar;  "Trustworthy IoT Data Streaming Using Blockchain and IPFS",   IEEE ACCESS,  2022.
9. Xinze Yu; Yunzhou Xie; Qiujie Xu; Zhuqing Xu; Runqun Xiong;  "Secure Data Sharing for Cross-domain Industrial IoT Based on Consortium Blockchain",   2023 26TH INTERNATIONAL CONFERENCE ON COMPUTER SUPPORTED ...,  2023.
10. Saleh Alghamdi; A. Albeshri; Ahmed Alhusayni;  "Enabling A Secure IoT Environment Using A Blockchain-Based Local-Global Consensus Manager",   ELECTRONICS,  2023.
11. Linchen Liu; Ruyan Liu; Zhiying Lv; Ding Huang; Xing Liu;  "Dual Blockchain-based Data Sharing Mechanism with Privacy Protection for Medical Internet of Things",   HELIYON,  2023.
12. Manzoor, A., Liyanage, M., Braeken, A., Kanhere, S. S., & Ylianttila, M. (2018). "Blockchain based Proxy Re-Encryption Scheme for Secure IoT Data Sharing." *arXiv preprint arXiv:1811.02276*
13. Zhang, Y., Kasahara, S., Shen, Y., Jiang, X., & Wan, J. (2021). "A Proxy Re-Encryption Approach to Secure Data Sharing in the Internet of Things Based on Blockchain." *IEEE Internet of Things Journal*, 8(2), 1150-1160.
14. Alharbi, A., & Aspinall, D. (2019). "A Secured Proxy-Based Data Sharing Module in IoT Environments Using Blockchain." *Sensors*, 19(5), 1235